# First integrals of Ginzburg-Landau equations and stability criteria for vortex-free state in unconventional superconductors

K.I. Kugel *, W.V. Pogosov **[1], A.L. Rakhmanov *

* Institute for Theoretical and Applied Electrodynamics, Russian Academy of Sciences, Izhorskaya ul. 13/19, Moscow, 127412 Russia
** Moscow Institute of Physics and Technology, Institutskii per. 7, Dolgoprydnyi, Moscow region, 141700 Russia

**Abstract**

The first integrals of the Ginzburg-Landau equations for a vortex-free state of superconductors with different mixed symmetries of the order parameter are found. The general boundary conditions for the order parameter at the ideal interface between the superconductor and vacuum are derived. Based on these integrals and boundary conditions, we analyze the stability criteria for vortex-free state in unconventional superconductors. The threshold field $H_s$ above which the Abrikosov vortices can enter the superconductor is found to be higher or equal to the thermodynamic critical field for all states under study.

**1. Introduction**

The symmetry of order parameter (OP) in high-temperature superconductors (HTSC) attracts a considerable current interest [1,2]. It is well known that the OP symmetry depends on crystal lattice symmetry, namely, superconducting states can be classified according to irreducible representations of the corresponding point group [3]. A lot of superconducting compounds have a tetragonal crystal lattice. The possible superconducting states in the tetragonal crystal belong to the following five representations $d_{x^2-y^2}$ (or $d$), $d_{xy(x^2-y^2)}$ (or $d_{xy}$), $s$, $s_{xy}$ and $E_{1g}$. The first four of them are one-dimensional, and so the OP is a single complex number, and $E_{1g}$ is the two-dimensional representational with two complex components of OP.

There is a considerable experimental evidence suggesting that the symmetry of HTSC is most likely to be $d$ wave, but some experiments give indication of the presence of an additional OP component [4,5]. It was shown in [6,7], that such a component always coexists with a predominantly $d$ wave OP in orthorhombic superconductors such as YBCO. Moreover, any inhomogeneity in the $d$ wave superconductor (for example, the surface) acts as a source of additional OP components. It is not yet clear which of the additional components really exists in HTSC. The two-component mixed states $d+d_{xy}$, $s+d$, and $d+s_{xy}$ are usually considered as possible kinds of actually existing OP symmetries.

In this paper, the first integrals of Ginzburg-Landau (GL) equations for these states are found. Starting from the most general symmetry considerations, the boundary conditions on ideal interface between the superconductor and vacuum are derived. As an example of

---

possible applications of the first integrals, we analyze the stability criteria for the vortex-free state in unconventional superconductors.

The flux lines begin to penetrate the type II superconductor having an ideal surface if the external magnetic field $H_e$ is equal to the thermodynamic critical field $H_c$ (Bean-Livingston surface barrier) [8,9]. However, the Bean-Livingston surface barrier is usually strongly suppressed in HTSC [10]. The measured penetration fields $H_s$ are much less than $H_c$ and usually of the order of the lower critical field $H_{c1}$. The first integrals of GL equations and the boundary conditions for OP allow us to find $H_s$ for the superconductors having different symmetries of OP. It is shown that in all cases this field is higher or equal to $H_c$. Thus, the nontrivial symmetry of OP can not be a cause for the suppression of the Bean-Livingston barrier in HTSC with the ideal surface.

**2. The first integrals of GL equations**

Let us consider the superconductor in the vortex-free state. The external field is assumed to be parallel to the sample surface. In this case, the magnetic field, supercurrent and OP vary along the direction normal to the surface. However, the problem is not purely one-dimensional since the magnetic field rotates in anisotropic superconductor. We study here the most general case of arbitrary angles between the vector $\mathbf{H}_e$ and the crystallographic axes.

The GL equations can be obtained by means of variation of the free energy functional with respect to OP and vector potential. Similarly, it is possible to obtain the boundary conditions for OP on the ideal surface if one takes into account additional free energy related to the surface. Thus, it is convenient to present the free energy functional in the form [11-13]

$$\Phi = \int F dV + \int f dS, \tag{1}$$

where $F$ is the bulk free energy density and $f$ is the surface one. The type of the boundary conditions depends on the superconducting state symmetry, since both $F$ and $f$ depend on it. Note, that in [11] the boundary conditions were found by this method for one-dimensional anisotropic OP and for two-dimensional $E_{1g}$ state.

First, let us study the state having mixed $d + d_{xy}$ OP symmetry. The OP in this case has two spatially varying complex components $\eta_l$, $l = 1,2$. In the framework of GL, theory the bulk free energy density can be written as [14]:

$$F = \sum_{l=1,2} \left( \alpha_l |\eta_l|^2 + \frac{\beta_l}{2} |\eta_l|^4 + \sum_{k=1}^{3} \frac{1}{2m_k^l} |\partial_k \eta_l|^2 \right) + \frac{\mathbf{H}^2}{8\pi} +$$
$$+ \tilde{k} \left( (\partial_1 \eta_1)^* (\partial_2 \eta_2) - (\partial_2 \eta_1)^* (\partial_1 \eta_2) + c.c. \right) + \gamma |\eta_1|^2 |\eta_2|^2 + \delta (\eta_1^2 \eta_2^{*2} + c.c.), \tag{2}$$

where $x_j$ ($j = 1,2,3$) are the crystal axes, $x_3$ is a four-fold axis, $\tilde{k}, \delta, \gamma, \alpha_l, \beta_l$ are the real coefficients, $\partial_j = -i\hbar \frac{\partial}{\partial x_j} - 2eA_j$ (**A** is the vector potential) and $m_k^l$ are the mass tensors for OP components in $x_j$ coordinate system. The corresponding GL equations are [14]:

$$\alpha_k \eta_k + \beta_k \eta_k |\eta_k|^2 + \gamma |\eta_{3-k}|^2 \eta_k + 2\delta \eta_{3-k}^2 \eta_k^* + \sum_{j=1}^{3} \frac{1}{2m_j^k} \partial_j^2 \eta_k + \quad (3)$$

$$+ 2ei\hbar \tilde{k}\, \eta_{3-k}(-1)^{k+1} H_3 = 0, \qquad k=1,2.$$

$$\frac{\nabla \times \nabla \times \mathbf{A}}{4\pi} = \mathbf{j} = \sum_{k=1,2} \sum_{j=1}^{3} \left( \frac{e\hbar}{im_j^k} \left( \eta_k^* \frac{\partial \eta_k}{\partial x_j} - \eta_k \frac{\partial \eta_k^*}{\partial x_j} \right) \mathbf{x}_j - 4e^2 |\eta_k|^2 \frac{A_j}{m_j^k} \mathbf{x}_j \right) + \quad (4)$$

$$+ 4\tilde{k} \nabla \times (\mathbf{x}_3 \, \mathrm{Im}\, \eta_1^* \eta_2),$$

where $H_3$ is the magnetic field component parallel to the four-fold axis $x_3$ and $\mathbf{x}_j$ are the unit vectors directed along the $x_j$ axes. According to [11,13], the additional phenomenological surface energy term for the two-component states can be presented in the following form:

$$f = \frac{b_1(\mathbf{n})}{2} |\eta_1|^2 + \frac{b_2(\mathbf{n})}{2} |\eta_2|^2 + b(\mathbf{n})(\eta_1 \eta_2^* + \eta_1^* \eta_2), \quad (5)$$

where $\mathbf{n}$ is the surface normal. The functions $b_1(\mathbf{n})$, $b_2(\mathbf{n})$, $b(\mathbf{n})$ depend on the angles between the surface normal and crystal axes and must be invariant under transformations from the tetragonal point group. The boundary conditions can be obtained by variation of the free energy functional with respect to OP. The surface contribution to the variation of the free energy can be presented as a sum of two surface integrals. These integrals arise from the surface free energy and from the gradient terms in the bulk free energy. The variation of (5) with respect to $\eta_1^*$ gives:

$$\oint \left( b_1(\mathbf{n}) \eta_1 + b(\mathbf{n}) \eta_2 \right) dS.$$

The surface integral derived from the bulk contribution to the free energy is:

$$-\oint \left( \sum_{j=1}^{3} n_j \frac{i\hbar}{2m_j^1} \partial_j \eta_1 + i\tilde{k}\hbar \left( n_2 \partial_1 \eta_2 - n_1 \partial_2 \eta_2 \right) \right) dS.$$

In order to obtain the boundary conditions, one should put the sum of these two integrals (that is, the surface contribution to the variation of the free energy) equal to zero. So the boundary conditions are:

$$\sum_{j=1}^{3} n_j \frac{\hbar}{2m_j^k} \partial_j \eta_k + \tilde{k}\hbar(-1)^{k+1} \left( n_2 \partial_1 \eta_{3-k} - n_1 \partial_2 \eta_{3-k} \right) =$$

$$= \frac{b_k(\mathbf{n})}{i} \eta_k + \frac{b(\mathbf{n})}{i} \eta_{3-k}, \qquad k=1,2 \quad (6)$$

For the sake of simplicity let us perform the gauge transformation as a result of which the phase of one of the OPs, for example $\eta_1$, will be constant inside the superconductor. It is convenient to choose it to be equal to zero. This procedure is possible in any simply connected sample in the absence of vortices.

Let us introduce the coordinate system with $z$ axis directed along the external field, $x$ and $y$ axes directed perpendicular and parallel to the surface, respectively. The GL equations for both OPs are complex and they can be written as four equations for real quantities:

$$\alpha_l \eta_l + \beta_l \eta_l^3 - k_l \eta_l'' + k_l \eta_l \varphi_l'^2 - 2e\hbar \eta_l \varphi_l' \sum_{j=1}^{3} \frac{A_j n_j}{m_j^l} + 2e^2 \eta_l \sum_{j=1}^{3} \frac{A_j^2}{m_j^l} - 2e\hbar \tilde{k} \eta_{3-l} (A_2' n_1 -$$

$$- A_1' n_2) \sin \varphi_2 + \gamma \eta_l \eta_{3-l}^2 + 2\delta \eta_l \eta_{3-l}^2 \cos 2\varphi_2 = 0; \quad l = 1,2 \tag{7}$$

$$-k_l \varphi_l'' \eta_l - 2k_l \eta_l' \varphi_l' + 2e\hbar \eta_l \sum_{j=1}^{3} \left( \frac{A_j n_j}{m_j^l} \right)' + e\hbar \eta_l \sum_{j=1}^{3} \left( \frac{A_j n_j}{m_j^l} \right)' -$$

$$- 2(-1)^l e\hbar \tilde{k} \eta_{3-l} \left( A_2' n_1 - A_1' n_2 \right) \cos \varphi_2 - 2(-1)^l \delta \eta_l \eta_{3-l}^2 \sin 2\varphi_2 = 0, \tag{8}$$

where all the derivatives are taken with respect to $x$, $\varphi_1$ is equal to zero and

$$k_l = \frac{\hbar^2}{2} \sum_{j=1}^{3} \frac{n_j^2}{m_j^l}, \quad l = 1,2.$$

The components of the vector potential satisfies the following equations:

$$-A_j'' + n_j (\nabla \mathbf{A})' = 8\pi e\hbar \varphi_2' \frac{n_j}{m_j^2} \eta_2^2 - 16\pi \sum_{i=1,2} \left( e^2 \eta_i^2 \frac{A_j}{m_j^i} \right) + 16 \tilde{k} (\eta_1 \eta_2 \sin \varphi_2)' p_j, \tag{9}$$

where vector $p_j = (n_2, -n_1, 0)$. Let us multiply (8) by $\eta_1$ at $l = 1$ and by $\eta_2$ at $l = 2$. The sum of the obtained equations can be presented as $I_1' = 0$, where

$$I_1 = \sum_{j=1}^{3} \frac{n_j \hbar^2}{2m_j^2} \eta_2^2 \left( \frac{\partial \varphi_2}{\partial x_j} - \frac{2e}{\hbar} A_j \right) - \sum_{j=1}^{3} \frac{n_j \hbar^2}{2m_j^1} \eta_1^2 \frac{2e}{\hbar} A_j. \tag{10}$$

We see that $I_1$ is the first integral of the GL equations.

To obtain one more first integral, we can use the following expression:

$$\frac{1}{8\pi} \left( \mathbf{H}^2 \right)' = \frac{1}{8\pi} \left[ 2A_1' A_1'' + 2A_2' A_2'' + 2A_3' A_3'' - \left( (div \mathbf{A})^2 \right)' \right]. \tag{11}$$

It can be transformed by substituting $A_j''$ from (9). Let us multiply (7) by $\eta_1'$ at $l = 1$ and by $\eta_2'$ at $l = 2$, and (8) by $(-\eta_2 \varphi_2')$ at $l = 2$. Summing these three equations and taking into account (11), one can find the next first integral $I_2$:

$$I_2 = \sum_{l=1,2} \left( \alpha_l |\eta_l|^2 + \frac{\beta_l}{2} |\eta_l|^4 - \sum_{j=1}^{3} \left( \frac{\hbar^2}{2m_j^l} \frac{\partial \eta_l}{\partial x_j} - 2e^2 \eta_l^2 \frac{A_j^2}{m_j^l} \right) \right) +$$

$$+ \gamma |\eta_1|^2 |\eta_2|^2 + \delta (\eta_1^2 \eta_2^{*2} + \eta_1^{*2} \eta_2^2) - \frac{\mathbf{H}^2}{8\pi}. \tag{12}$$

The method of obtaining of the first integrals and the boundary conditions for the $s+d$, $d+s_{xy}$ states is similar to that in the $d+d_{xy}$ case. For each of these symmetries we summarize below the first integrals and the boundary conditions.

1) $s+d$.

The free energy expansion differs from the case of $d+d_{xy}$ symmetry by the coupled gradient terms, which can be written as [14]:

$$\tilde{k}\left((\partial_1\eta_1)^*(\partial_1\eta_2) - (\partial_2\eta_1)^*(\partial_2\eta_2) + c.c.\right).$$

The boundary conditions are:

$$\sum_{j=1}^{3} n_j \frac{1}{2m_j^k} \partial_j \eta_k + \tilde{k}\left(n_1\partial_1\eta_{3-k} - n_2\partial_2\eta_{3-k}\right) = \frac{b_k}{i}\eta_k + \frac{b}{i}\eta_{3-k}, \quad k=1,2. \qquad (13)$$

The first integrals are:

$$I_1 = \sum_{j=1}^{3} \frac{n_j\hbar^2}{2m_j^2} \eta_2^2 \left(\frac{\partial \varphi_2}{\partial x_j} - \frac{2e}{\hbar} A_j\right) - \sum_{j=1}^{3} \frac{n_j\hbar^2}{2m_j^1} \eta_1^2 \frac{2e}{\hbar} A_j + $$

$$+ \frac{\tilde{k}}{2i}\left(n_2\left(\eta_2^*\partial_2\eta_1 - \eta_1^*\partial_2\eta_2\right) - n_1\left(\eta_2^*\partial_1\eta_1 - \eta_1^*\partial_1\eta_2\right)\right) \qquad (14)$$

$$I_2 = \sum_{l=1,2}\left(\alpha_l|\eta_l|^2 + \frac{\beta_l}{2}|\eta_l|^4 - \sum_{j=1}^{3}\left(\frac{\hbar^2}{2m_j^l}\left|\frac{\partial \eta_l}{\partial x_j}\right|^2 - 2e^2|\eta_l|^2\frac{A_j^2}{m_j^l}\right)\right) + $$

$$+ 4\tilde{k}e^2\left(\eta_1\eta_2^* + k.c\right) + \hbar^2\tilde{k}\left(\frac{\partial \eta_1^*}{\partial x_1}\frac{\partial \eta_2}{\partial x_1} - \frac{\partial \eta_1^*}{\partial x_2}\frac{\partial \eta_2}{\partial x_2} + c.c.\right) + \qquad (15)$$

$$+ \gamma|\eta_1|^2|\eta_2|^2 + \delta\left(\eta_1^2\eta_2^{*2} + \eta_1^{*2}\eta_2^2\right) - \frac{\mathbf{H}^2}{8\pi}.$$

2) $d+s_{xy}$.

The characteristic feature of this state is the absence of the coupled gradient terms in the free energy expansion [14]. So, this state can be regarded as the special case of $d+d_{xy}$ or $s+d$ at $\tilde{k}=0$.

## 3. Surface barrier

It is well known that the Abrikosov vortices interact with the surface. This interaction leads to the Bean-Livingston surface barrier, which impede the penetration of the flux lines into a superconductor [8]. As a result, in the increasing magnetic field the vortices begin to penetrate the sample when the external field is equal to a certain value $H_s$ higher than the first critical field $H_{c1}$. De Gennes obtained that in the framework of GL theory the field of the vortex penetration into a bulk isotropic superconductor with the ideal surface is equal to the thermodynamic critical field $H_c$ [9].

In this section, we shall analyze the stability criteria for the vortex-free state in the external magnetic field. We shall consider the cases of the unusual superconducting states for which the first integrals were found in Section 2. To outline the proposed approach, we consider at first the usual isotropic superconductor. This case was studied by de Gennes for the superconductors with GL parameter $\kappa \gg 1$ [9]. In [9] OP is assumed to satisfy the boundary condition $\eta'=0$, standard for the GL theory. We modify the method used by de Gennes for more general boundary condition and arbitrary values of $\kappa$.

### 3.1. *Isotropic superconductors*

Let us consider a semi-infinite superconductor in the external magnetic field $H_e$. For convenience we use dimensionless quantities (all the distances are normalized by the London penetration depth, OP is normalized by its equilibrium value far from the surface in the absence of the external field, and the magnetic field is normalized by $H_c/\sqrt{2}$). The expressions for the first integral $I_2$ and the boundary condition (6) for OP can be written as

$$I_2 = \frac{1}{2}\eta^4 - \eta^2 - \frac{1}{\kappa^2}\eta'^2 + a^2\eta^2 - h^2 \tag{16}$$

$$\eta'(0) = \eta(0)/l, \tag{17}$$

where $\kappa$ is the GL parameter, which is equal to the ratio of London penetration depth and the coherence length, $a$ and $h$ are the dimensionless vector potential and magnetic field, and $l$ is the so called "extrapolation length". Far from the surface the supercurrent and the magnetic field equal zero and, therefore, $\eta(\infty) = 1$ and $a(\infty) = 0$. So, one can find the value of the first integral: $I_2 = -1/2$.

Using (16) and (17), one can obtain the following relationship between the dimensionless external field $h_e$ and the value of OP at the superconductor surface $\eta(0)$

$$h_e^2 - \frac{1}{2} = \frac{1}{2}\eta(0)^4 - \eta(0)^2 \frac{1}{\kappa^2 l^2} + a(0)^2\eta(0)^2 - \eta^2(0). \tag{18}$$

This equation implies, that the value $\eta(0)$ depends on the external field $h_e$. From the definition $\eta(0)$ can vary from zero to one. Thus, the maximum external field $h_s$ which does not destroy the vortex-free state (the homogeneous state in the plane *yz*) can be defined as the maximum field $h_e$, which satisfies (18) under the condition that $\eta(0)$ changes from 0 to 1. It is quite a complicated task to find the maximum of right-hand side of equation (18). In general case the dependence of vector potential *a* on the order parameter $\eta$ at the surface can be found only by the numerical solution of the complete set of Ginzburg-Landau equation. However it is easy to see, that there exists a particular solution: $h_e = 1/\sqrt{2}$ at $\eta(0) = 0$. We defined above $h_s$ as the maximum field meeting (18). Hence, we have $h_s \geq 1/\sqrt{2}$ or $h_s \geq h_c$.

Let us consider the case when $\kappa \gg 1$, which is typical for HTSC. The GL equations for OP can be written as

$$-\kappa^{-2}\eta'' + \eta^3 - \eta + a^2\eta = 0. \tag{19}$$

Since $\kappa \gg 1$, one can neglect $-\kappa^{-2}\eta''$ in (19) and substitute the expression for $a^2$ from this equation to the first integral (16). In this case, the dependence of $\eta(0)$ on the external field is defined by the equation:

$$h_e^2 - h_c^2 = -\frac{1}{2}\eta(0)^4 - \eta(0)^2 \frac{1}{l^2\kappa^2}. \tag{20}$$

It follows from (20) that the maximum value of $h_e$ is achieved when OP is equal to zero on the surface, and we get that $h_s = h_c$.

Thus, in the isotropic case, the vortex-free state in the sample with an ideal surface can exist when external field $H_e$ is less than certain value $H_s$ that exceeds or equals the thermodynamic critical field and $H_s = H_c$ at $\kappa \gg 1$.

The method used in this subsection can be applied in the cases of more complex symmetries. The knowledge of the first integrals significantly facilitates the solution.

### 3.2. *Unconventional superconductors*

Let us study the case of $d+d_{xy}$ symmetry. Far from the surface, the supercurrent and the magnetic field are equal to zero and therefore the values of the first integrals are

$$I_1 = 0, \; I_2 = -\frac{H_c^2}{8\pi}. \tag{21}$$

Using expression (12) for the first integral $I_2$ and taking into account conditions (21) it is possible to obtain the following equation, which relates the external field to the values of $\eta_1, \eta_2, \eta_1'$ and $\eta_2'$ at the surface

$$\sum_{l=1,2}\left(\alpha_l\eta_l^4(0)+\frac{\beta_l}{2}\eta_l^4(0)-k_l\eta_l'^2(0)+2e^2\eta_l^2(0)\sum_{j=1}^3\frac{A_j^2(0)}{m_j^l}\right)-\frac{\mathbf{H}_e^2}{8\pi}$$
$$+k_2\eta_2^2(0)\varphi_2'^2(0)+\gamma\,\eta_1^2(0)\eta_2^2(0)+2\delta\,\eta_1^2(0)\eta_2^2(0)\cos2\varphi_2(0)=-\frac{\mathbf{H}_c^2}{8\pi}. \tag{22}$$

Similar to the isotropic case, we can also find here the particular solution of (22). The relationship between $\eta_1'(0), \eta_2'(0), \varphi_2(0),$ and $\varphi_2'(0)$ follows from the boundary conditions (6) and the value of the first integral $I_1$ (21). The relationship between $\eta_1(0)$ and $\eta_2(0)$ can be found from the solution of GL equations (7)-(9). However, at $\kappa\gg1$ it is possible to neglect $k_1\eta_1''$ in equation (7) at $l = 1$, similarly to the isotropic superconductor. For $\eta_1(0)=0$, it follows from (7), that $\eta_2(0)$ =0. For this values of $\eta_1(0)$ and $\eta_2(0)$, we get from (22) that $H_e = H_c$. From the definition of $H_s$ as the maximum external field, which does not destroy the vortex-free state, we have $H_s \geq H_c$.

Similarly to the $d+d_{xy}$ case, it is possible to find that the penetration field is not smaller than the thermodynamic critical field also for $s+d$ and $d+s_{xy}$ symmetries.

Let us now consider the case of anisotropic one-component OP. The first integrals can be determined using the integral for the two-component OP and putting one of the components equal to zero. It is possible to reduce this problem to the case of isotropic superconductor by passing to new dimensionless variables. The distances along each of the principal axes are normalized by the corresponding London penetration depth and the Ginzburg-Landau parameter should be changed by its effective value $\tilde{\kappa}$, defined as:

$$\tilde{\kappa}^{-2} = n_1^2\kappa_1^{-2} + n_2^2\kappa_2^{-2} + n_3^2\kappa_3^{-2}, \tag{23}$$

where $\kappa_i$ are the values of Ginzburg-Landau parameter along the principal axes.

The symmetry of one-component *d*-wave OP implies that the extrapolation length *b* in boundary condition (17) attains its minimum at $n_1 = n_2 = 1/\sqrt{2}$, $n_3 = 0$ ([110] surface) [11]. The microscopic treatment shows that *b* in this case can be smaller than the coherence length [15]. The small value of extrapolation length means that OP is significantly suppressed at [110] surface. In this sense, the boundary condition *f* = 0 is sometimes introduced [11,16]. Actually, the situation with $b \to 0$ reduces to our previous treatment. Although, to verify our results, we solved numerically the complete set of the Ginzburg-Landau equations for the case *f* = 0 at the surface. The use of first integral (16) allowed us to reduce the number of independent variables to a single one. This made the problem much simpler. The numerical calculations also lead to condition $H_s = H_c$ for all $\kappa > 1$.

## 4. Conclusion

In this paper, we found the first integrals of the Ginzburg-Landau equations for the superconducting states with mixed symmetries of the order parameter. The boundary conditions for the order parameter at the surface between the superconductor and vacuum were derived for $d + d_{xy}$, $s + d$, and $d + s_{xy}$ symmetries in a rather general case. The first integrals and the boundary conditions were applied for the analysis of the stability criteria for the vortex-free state in the external magnetic field. The maximum external magnetic field, which does not destroy the vortex-free state, was shown to be higher or equal to the thermodynamic critical field. Hence, the unusual symmetry of the order parameter can not be the cause of the suppression of the Bean-Livingston surface barrier in HTSC.

**Acknowledgments**

The work was supported by the Russian Foundation for Basic Research (RFBR), grants #00-02-18032 and #00-15-96570, by the joint INTAS-RFBR program, grant #IR-97-1394, and by the Russian State Program 'Fundamental Problems in Condensed Matter Physics'.